# How creative versus technical constraints affect individual learning in an online innovation community

**Victor P. Seidel**
FW Olin Graduate School of Business
Babson College
vseidel@babson.edu

**Christoph Riedl**
D'Amore-McKim School of Business
Northeastern University
c.riedl@northeastern.edu

*Manuscript of March 25, 2023*

## ABSTRACT

Online innovation communities allow for a search for novel solutions within a design space bounded by constraints. Past research has focused on the effect of creative constraints on individual projects, but less is known about how constraints affect learning from repeated design submissions and the effect of the technical constraints that are integral to online platforms. How do creative versus technical constraints affect individual learning in exploring a design space in online communities? We analyzed ten years of data from an online innovation community that crowdsourced 136,989 design submissions from 33,813 individuals. We leveraged data from two types of design contests—creatively constrained and unconstrained—running in parallel on the platform, and we evaluated a natural experiment where a platform change reduced technical constraints. We find that creative constraints lead to high rates of learning only if technical constraints are sufficiently relaxed. Our findings have implications for the management of creative design work and the downstream effects of the technical constraints of the information systems that support online innovation communities.

*Keywords:  Creative design; Constraints; Learning; Innovation Processes; Online Communities; Crowdsourcing*

# 1. Introduction

Online innovation communities provide the ability to engage large numbers of individuals in the search for new solutions (Faraj et al. 2016), and one form, online design communities, can be used to provide firms with expanded sources of creative insights (Majchrzak & Malhotra, 2016). Creative design is an important component of innovation, as successful designs help products gain acceptance (Hargadon & Douglas, 2001) and provide aesthetic value (Rindova & Petkova, 2007). Creative design can be conceptualized as the search for novel solutions within a "design space" of possibilities (MacLean et al., 1991; Sinha & May, 1996). Design spaces are conceptual spaces of possibilities co-created between designers and their audience (Biskjaer et al., 2014; Boden, 2004) and are defined by constraints of the bounds of possible solutions (Beaudouin-Lafon & Mackay, 2009). One central decision in using online communities for sourcing new designs is whether participants should be given broad or narrow constraints on their designs. Our study focus is on how constraints affect learning in online innovation communities.

The dominant model of the role of constraints implies that there is an overall inverted-U relationship between constraints and creative outcomes (Acar et al., 2019; Caniëls & Rietzschel, 2015). The advantages of constraints come from spurring on creative efforts and focusing attention (Chua & Iyengar, 2008; Stokes, 2005) while negative effects arise due to fixating on narrower solutions (Bayus, 2013; Jansson & Smith, 1991; Smith, 2003) or decreasing motivation (Amabile, 1983; Smith, 2003). Thus, a moderate amount of constraints results in most creative outcomes. While past research establishing this consensus model has focused on creative performance in singular design projects (e.g. Rosso, 2014) much less is known on the effect of constraints on learning, when individuals repeatedly explore a design space. As learning is a central aspect of how online communities are sustained (Lakhani & Wolf, 2005), and learning can provide advantages to both community members and their platform sponsors (Jin et al., 2021; Riedl & Seidel, 2018), a lack of focus on learning constitutes a significant gap in our understanding of the role of constraints. When individuals explore a design space repeatedly in order to learn how to make better contributions, they will be subject to two types of constraint common in creative work (Stokes, 2005): Not only the creative constraints of what designs are valued by the community but also technical constraints that are embedded in the affordances of the medium, such as an online platform. In this Research Note, our specific research question is: How do creative versus technical constraints affect individual learning in exploring a design space in online communities?

We analyzed ten years of data from an online innovation community hosting repeated creative design competitions comprising 136,989 design submissions by 33,813 individual members. We examined how creative and technical constraints affected how individuals learned to improve their performance in providing solutions as they explored a design space. We apply panel regression coupled with a natural experiment of changed technical constraints to disentangle the role creative versus technical constraints on individual learning. We find that creative constraints lead to high rates of learning only if technical constraints are sufficiently relaxed. We contribute a more nuanced theory on the role of different constraints as they relate to innovation and learning, insights into the role of technical constraints on creative work, and findings that have implications for the management of creative work in both online and offline settings.

## 2. Theoretical background and hypotheses

Online innovation communities allow for many individuals to share knowledge of possible solutions to a host platform, and the type of platform and community can take different forms depending on the context of the innovation challenge (Majchrzak & Malhotra, 2016). Some platforms connect to individuals who may have knowledge of a technological solution to a posted challenge, such as in a crowdsourced innovation contest (Jeppesen & Lakhani, 2010). In such technological innovation contexts, one may be able to objectively evaluate potential solutions to a given challenge and select the best solution of those offered. A different form of innovation community is focused on creative design, and these communities are defined by a different form of evaluation. Creative design projects are characterized by subjective audience evaluation of outcomes that provide designers with a changing and ongoing understanding of what is valued in a design space (Biskjaer et al., 2014; Fraiberger et al., 2018). Capturing the subjective nature of design, Stefano Giovannoni, a celebrated designer for the Italian firm Alessi, related: "Objects are not beautiful or ugly but are either suited or not to their time" (Fiell & Fiell, 2005,p. 95). The exploration of a creative design space is therefore a process of gaining an understanding of what is valued by such an audience at a certain moment in time—through evaluation signals such as written feedback or scoring—as the designer and audience together co-create understandings of value as new possibilities are generated and received (Boden, 2004). Over time, individuals in an online design community can learn how to improve their performance to meet the standards of those who are evaluating their designs (Riedl & Seidel, 2018).

Constraints are central to design and innovation, as a design space by its nature "constrains design possibilities on some dimensions, while leaving others open for creative exploration" (Beaudouin-Lafon & Mackay, 2009, p. 9). In a creative design endeavor two constraints are prominent: There are creative constraints, such as themes or genres to incorporate, and there are technical constraints, such as image resolution or colors available (Acar et al., 2019; Stokes, 2005). For example, in the realm of video game design there may be creative constraints on the genre of game (such as a "child-friendly non-violent adventure game"), as well as technical constraints (such as specific game console platform with video resolution limits). The following section develops hypotheses that consider the the learning effects on individuals that make repeated submissions within a design space, under creative versus technical constraints.

## 2.1. Effect of creative constraints on learning

Prior extensive research on constraints has developed a dominant model whereby constraints are theorized to be helpful for the outcome of a creative project, under an inverted-U relationship (Acar et al., 2019; Kornish & Hutchison-Krupat, 2017). An absence of constraints in an innovation task—such as a creative designer with merely a "blank page" before them—can be challenging due to an inability to choose direction (Haas et al., 2014). In contrast, the introduction of constraints and narrowing of options can generatively focus attention (Stokes, 2005) and free oneself from being overwhelmed with choice (Chua & Iyengar, 2008). However, too many constraints can lead to fixation effects on narrower solutions (Bayus, 2013; Jansson & Smith, 1991; Smith, 2003) and decreased motivation (Amabile, 1983; Smith, 2003).

Because learning plays a central role in the potential benefits of online innovation communities, our focus in this Research Note is on the effect of constraints on *learning* how to improve creative outcomes over repeated submissions. Improving performance through repeated design submissions will be determined by the creative skills developed over time—such as how to develop a new realistic character in a video game or how to create an attractive case for a new generation of mobile phone. In domains such as architecture and music, a series of imposed constraints across a variety of tasks has been shown to lead to improving a repertoire of artistic skills (Stokes, 2005) that may help not only a singular project but also learning in how to use such skills to gain positive evaluations in a design space over repeated project submissions.

Individual learning is based on making sense of the feedback signals from individual efforts (Argote, 2012; Newell & Rosenbloom, 1981). If evaluation signals are hard to interpret, then learning is impeded (Bandura, 1977), and so any effort to make such signals less ambiguous should advance learning. Creative designs are evaluated based on the aesthetics of the design (how well it was executed) as well as how desirable the design is to the audience (the subject). Feedback on design submissions can often be challenging to parse out how one is being evaluated on aesthetics versus desirability, and so creative constraints should help to provide clarity about what is being evaluated. Work on crowdsourcing contests has shown that high quality knowledge sharing is required the improve the performance of other members (Jin et al., 2021). Under higher constraints about the creative task at hand, feedback will be easier to decipher with regard to the skill by which one created aesthetic value, leading to increased capacity to learn from such feedback. For example, feedback when exploring a relatively unconstrained design space of "new lamps for the home" will be harder to decipher than feedback constrained to a design space of "new small desk lamps for a home office." Working on a succession of creatively constrained submissions should lead to faster learning than just submitting to unconstrained design challenges. Therefore, we hypothesize:

*H1: Engaging in a higher proportion of submissions with creative constraints will accelerate performance improvement (higher rate of learning)*

## 2.2. The effect of technical constraints

In any creative endeavor there are both imposed constraints (decided upon for the task at-hand, such as creative constraints in design) and implicit constraints determined by the context (Elster, 2000). For example, an implicit constraint for a video game designer may be the speed of a console's processor, thereby constraining the visual details of characters and scenes that need to be rendered on screen in real time. Such implicit technical constraints will reduce the range of experimentation possible within a creative design space (Beaudouin-Lafon & Mackay, 2009). Limited range of experimentation has the effect of decreasing the variance not only of possible creative outcomes but also of signals that one can use to update an understanding of what is valued by an audience (Bandura, 1977). Furthermore, when faced with technical constraints, a designer may need to focus their attention on working around such constraints, rather than focusing on their creative design, and the management of attention is critical for problem-solving in online communities (Haas et al., 2014). Together these effects—limited range of experimentation, limited range of feedback, and shifted attention—are then likely to have negative consequence on the rate of learning for designers who experience increased technical constraints.

In contrast, relaxing technical constraints will allow a broader range of experimentation, and such experimentation will provide the designer with more variation in feedback which helps a designer to better understand how designs are accepted by an audience. Therefore, we hypothesize:

*H2. Creating design submissions under decreased technical constraints will accelerate performance improvement (higher rate of learning)*

While our hypotheses are developed agnostic to the research setting in which design takes place, the rich digital trace data of online communities provide the ability to precisely measure the effect of constraints on creative design projects (Ehls et al., 2020; Erat & Kavadias, 2008). Furthermore, such communities are also the context where such work is increasingly done and where individuals repeatedly explore the same design space. For our study of the learning effects under the repeated exploration of a design space, we drew upon an online setting that would afford us a wealth of data by which to measure the impact of creative and technical constraints on performance improvement.

## 3. Research method

We used a ten year panel of T-shirt designs submitted to the Threadless website (www.threadless.com), an online platform for creative design competitions. At the time of the study Threadless supported two types of competitions: "regular competitions" were held every week and were creatively unconstrained, with only standard restrictions against such things as plagiarism and profanity. In contrast, "themed competitions" had specific creative constraints, with examples that included challenges to depict themes related to cubism, to reflect styles from the 1990s, and to address "circus" themes. Across the ten years there were data from 33,813 designers making 136,989 design submissions. For each design submitted, an audience of fellow designers and potential customers would evaluate the design by voting on a five-point scale, and there were 149 million audience votes cast in our data. Figure 1 provides an overview of how designers in our dataset chose to submit between unconstrained regular competitions (the x-axis) and constrained themed competitions (the y-axis), plotting their cumulative experience across the two types of competitions. While some designers appear to focus on either one or the other, most gain experience by working on a mix of unconstrained and constrained design projects.

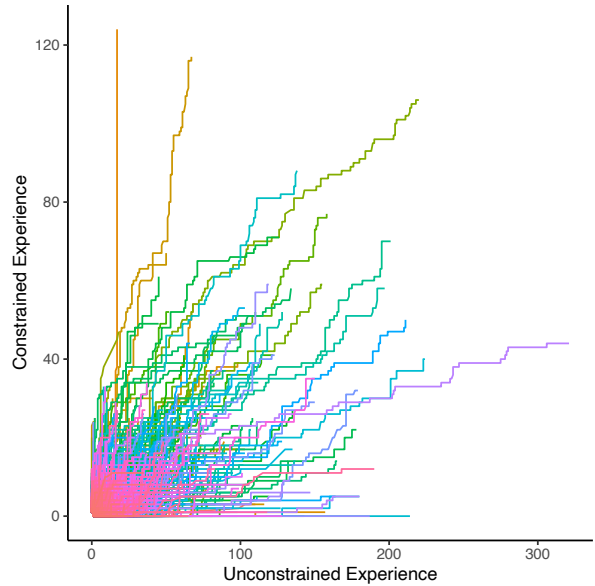

**Figure 1: Plot of designer experience in creatively unconstrained and creatively constrained submissions.**
Figure notes: Each designer's cumulative experience submitting to creatively unconstrained (x-axis) and creatively constrained (y-axis) contests is plotted. Note that axis scales are uneven to account for relative number of each type of contest.

*Dependent Variable: Score.* The dependent variable *Score* measures the audience evaluation of a design submitted. *Score* is the mean of all audience votes in a the online community a design received. Overall, the dependent variable is the aggregate of a total of 149 million votes with an average of 830 votes per design submission (SD=651.1). In our empirical setting "learning" is reflected in an increasing score over time.

*Constrained and unconstrained experience.* Prior experience from directly engaging in design projects from either constrained or unconstrained contests are measured prior to submitting the current design ($Exp_{it-1}$). Each contest is either a constrained or unconstrained contest. We measure the proportion of constrained ("themed competitions") to understand the role of such experience in promoting learning how to develop highly-evaluated designs.

*Control Variables.* We include a control variable *Tenure*, which captures the number of years a designer has been active on the Threadless platform. *Tenure* effectively also captures the general passage of time and thus incorporates possible experience accumulated outside the organization through training, education, or from other design activity. We also include an indicator *Past Winner*, which equals one (1) if a designer of a submission has been successful in the past (i.e., has been printed) and zero (0) otherwise. This measure thus controls for increased popularity of a known and successful designer, which may bias community voting upward. In a similar vein, we include *Vote Count* to capture the number of ratings that a submission has received (and thus make up Score) as more popular designs tend to have higher scores. Table 1 shows summary statistics and correlations of our key variables.

**Table 1: Descriptive statistics and correlations of key variables**

|  | Mean | SD | Min | Max | (1) | (2) | (3) | (4) | (5) | (6) | (7) | (8) | (9) |
|---|---|---|---|---|---|---|---|---|---|---|---|---|---|
| $Score_{it}$ (1) | 1.91 | 0.51 | 0.25 | 5.00 |  |  |  |  |  |  |  |  |  |
| $ScoreCount_{it}$ (2) | 850.61 | 657.46 | 1.00 | 3791.00 | 0.30 |  |  |  |  |  |  |  |  |
| $Unconstrained_t$ (3) | 0.85 | 0.36 | 0.00 | 1.00 | -0.06 | -0.07 |  |  |  |  |  |  |  |
| $Exp_{it-1}$ (4) | 16.47 | 30.84 | 1.00 | 365.00 | 0.31 | 0.06 | -0.06 |  |  |  |  |  |  |
| $Exp(Unconstrained)_{it-1}$ (5) | 13.65 | 25.09 | 0.00 | 321.00 | 0.30 | 0.05 | -0.02 | 0.98 |  |  |  |  |  |
| $Exp(Constrained)_{it-1}$ (6) | 2.83 | 8.06 | 0.00 | 124.00 | 0.27 | 0.07 | -0.17 | 0.78 | 0.63 |  |  |  |  |
| $Votes_{it-1}$ (7) | 2080.83 | 6498.96 | 0.00 | 128864.00 | 0.24 | 0.05 | -0.08 | 0.44 | 0.41 | 0.41 |  |  |  |
| $Tenure_{it}$ (8) | 0.83 | 1.13 | 0.00 | 8.84 | 0.30 | 0.02 | -0.12 | 0.44 | 0.40 | 0.45 | 0.37 |  |  |
| $Pastwinner_{it}$ (9) | 0.11 | 0.32 | 0.00 | 1.00 | 0.38 | 0.14 | -0.08 | 0.56 | 0.52 | 0.52 | 0.39 | 0.45 |  |
| $AfterColorChange_{it}$ (10) | 0.55 | 0.50 | 0.00 | 1.00 | 0.24 | -0.32 | -0.04 | 0.18 | 0.17 | 0.14 | 0.13 | 0.25 | 0.11 |

Table note: All pairwise correlations are statistically significant at *p* < 0.001

Next we explain our empirical model. We follow the common approach to study learning (e.g., Argote and Epple 1990, Wiersma 2007) and estimate the rate of performance improvement (i.e., learning rate) from direct and indirect experience over time. A baseline model is given in Equation (1):

$$\ln Score_{it} = \beta_1 \ln Exp_{it-1} + \beta_2 \ln Votes_{it-1} + \alpha_i + \epsilon_{it}. \tag{1}$$

Here, $Score_{it}$ is the performance of the *t*th submission of individual *i*. $Exp_{it-1}$ is the cumulative prior direct experience (number of prior submissions) and $Votes_{it-t}$ is the cumulative prior rating experience, $\alpha_i$ are individual-level random effects, $\alpha_t$ are contest-level random effects, and $\epsilon_{it}$ are are error terms. By taking the log of the terms we are able to capture the *rate* of change and thus capture learning. To test whether the source of experience—coming from either constrained or unconstrained contests matters— we include the proportion of "units of experience" coming from constrained contests. Finally, to test if the experience from constrained contests spills over to unconstrained contests, we include an interaction term between the proportion of experience and a dummy indicating whether the contest at time *t* was unconstrained. The main effect of the dummy drops out as it is colinear with the contest fixed effect. The equation of our full model is

$$\begin{aligned} \ln Score_{it} = & \beta_1 \ln Exp_{it-1} + \beta_2 \ln Votes_{it-1} \\ & + \beta_3 ConstrainedExp_{it-1} \\ & + \beta_4 ConstrainedExp_{it-1} \times UnconstrainedContest_t \\ & + \beta_{5-7} Controls_{it} + \beta_{8-612} Contest_t + \alpha_i + \epsilon_{it}. \end{aligned} \qquad (2)$$

Faster learning from constrained experience is supported if $\beta_3$ is positive and spillovers are supported if $\beta_4$ is positive.

Given the crossed panel nature of our data, we adopt a multilevel estimation approach (Gelman and Hill, 2006) to accommodate the nesting of repeated measures within designers and competitions, and accounting for autocorrelation and unequal spacing of observations in time. We specify contests as fixed effects and individuals as random effects. When the levels in a dataset represent all possible levels of the factor about which inferences are made, fixed effects are most appropriate. In our setting, this is the case for contests as our data includes all contests that took place during the 10-year period. A fixed-effects formulation assumes that estimating different intercept terms for each unit can best capture differences across units. This approach is appropriate if units differ in their average level of the outcome measure, which is likely in our case as contests vary in popularity, difficulty, and prize money, which likely affects the quality of submissions. A random-effects specification, on the other hand, assumes that levels of the factor used in the study represent a random sample of a larger set of potential levels (Greene 2003). Since we wish to generalize the learning among individuals in our sample to participants in crowdsourcing contests more generally, we specify random effects for individuals. We estimate the model using linear mixed model fit by maximum likelihood. We used the lme4 package (Bates et al., 2015) available in R (R Core Team, 2015). This form of multilevel modeling is the recommended approach for hierarchical data with multiple nested and crossed levels (Gelman & Hill, 2006). Furthermore, we can flexibly deal with varying numbers of observations per individual and contest. We performed several specification checks, including specifying the contest level as random instead of fixed, and the individual level as fixed instead of random. Results are consistent across different combinations.

*Natural Experiment.* In the seventh year of the ten years of data there was a change in which the technical constraint on the number of colors in the design palette was decreased, forming a natural experiment of the role of technical constraints. While at first only five colors could be used, this constraint was relaxed to eight colors. This 60% increase in the color palette provided a means to test how an increase in the possible design space for experimentation might change the degree of learning by designers. We analyze this natural experiment using standard difference-in-difference approach for exogenous shocks from quasi experiments (Goldfarb et al., 2022). The dummy variable After equals 1 if the contest *t* occurs after the technical constraints were relaxed, and 0 otherwise. Its main effect is absorbed by the contest fixed effects. In our main analysis, we use a 6-months time window before and after the platform change and look at longer time windows as robustness tests. Table 2 shows the number of observations in the narrow six month time window.

|        | Control | Treatment |
|--------|---------|-----------|
| Before | 7,742   | 1,508     |
| After  | 8,376   | 1,193     |

**Table 2: Observations in the natural experiment**

## 4. Results

Our first hypothesis was that submitting to a higher proportion of creatively constrained projects would accelerate learning how to produce designs that gain high audience evaluations. Table 3 shows the estimation models from our panel regression. The effect of the proportion of constrained project past experience (*Constrained Experience %tile*) is positive and significant ($\beta=0.045^{***}$), supporting H2. Model 2 also looks at the interaction effect (*Constrained Experience %ile x Unconstrained Competition*) to confirm that this effect holds for unconstrained projects as well as those for constrained projects ($\beta=0.015^{**}$). Overall, working on creatively constrained projects accelerated learning how to increase evaluation scores in the design space for both constrained and unconstrained design projects.

**Table 3: Model results of effect of constrained design experience on learning**

| Dependent Variable: | Learning (lnScore) | |
|---|---|---|
| | Source of Experience | Interaction with Unconstrained Contest |
| | (1) | (2) |
| Intercept | 1.091*** | 0.898*** |
| | (0.133) | (0.067) |
| lnExp | 0.002* | 0.002* |
| | (0.001) | (0.001) |
| Constrained Experience (%ile) | 0.045*** | 0.040*** |
| | (0.006) | (0.007) |
| lnVotes | 0.010*** | 0.010*** |
| | (0.000) | (0.000) |
| Tenure | 0.007*** | 0.007*** |
| | (0.001) | (0.001) |
| PastWinner | 0.029*** | 0.029*** |
| | (0.003) | (0.003) |
| ScoreCount | 0.199*** | 0.199*** |
| | (0.001) | (0.001) |
| Constrained Experience (%ile) × Unconstrained Competition | | 0.015** |
| | | (0.006) |
| Contest | Fixed | Fixed |
| Individual | Random | Random |
| AIC | −75368.654 | −75363.886 |
| Log Likelihood | 38296.327 | 38294.943 |
| Num. obs. | 136,989 | 136,989 |
| Variance: | | |
| Var: UserID (Intercept) | 0.013 | 0.013 |
| Var: Residual | 0.027 | 0.027 |

***$p < 0.001$, **$p < 0.01$, *$p < 0.05$

Our second hypothesis addressed the effect of decreasing technical constraints, expecting relaxed constraints to further improve learning. We used a difference-in-difference regression, comparing the learning rates in a window before and after the change in number of colors allowed, as shown in Table 4. Interestingly, the increased learning effect from creative constraints is only significant after the technical constrain (number of colors) is relaxed ($\beta = 0.047; p < 0.05$). This result modifies our expectation with respect to H2 that looked at all years; if there are too many technical constraints, it is not possible to increase the learning rate by doing more creatively constrained projects. We perform several robustness test. First, we perform a placebo test (Goldfarb, Tucker, and Wang 2014) in which we add a "fake" after date for a point in time at which no platform change occurred and we thus would expect to see no significant effect (four months after the technical constraints on colors were dropped). The interaction with the Fake After dummy is not significant (Model 2: $\beta = -0.037; NS$) while the interaction with After increases slightly in both size and significant level ($\beta = 0.058; p < 0.01$). Second, we altered our +/- 6 month time window in considering the effect of removing the constraint to include more design submissions (Model 3: 1 Year; Model 4: 2 Years; Model 5: and 2 year windows, as well as the entire dataset) but found the same results.

**Table 4: Difference-in-difference regression of natural experiment testing for the effect of relaxing technical constraints**

| Dependent Variable: | Learning (lnScore) | | | | |
|---|---|---|---|---|---|
| | ± 6 Months | Placebo Test | ± 1 Year | ± 2 Years | All Time |
| | (1) | (2) | (3) | (4) | (5) |
| (Intercept) | 0.268*** | 0.268*** | 0.279*** | 0.574*** | 1.120*** |
| | (0.019) | (0.019) | (0.020) | (0.018) | (0.133) |
| lnExp | 0.006*** | 0.006*** | 0.005*** | 0.004*** | 0.002** |
| | (0.001) | (0.001) | (0.001) | (0.001) | (0.001) |
| Constrained Experience (%ile) | 0.014 | 0.014 | −0.000 | −0.003 | −0.002 |
| | (0.016) | (0.016) | (0.012) | (0.010) | (0.009) |
| lnVotes | 0.006*** | 0.006*** | 0.006*** | 0.009*** | 0.010*** |
| | (0.001) | (0.001) | (0.000) | (0.000) | (0.000) |
| Tenure | 0.000 | 0.000 | 0.008*** | 0.014*** | 0.006*** |
| | (0.002) | (0.002) | (0.001) | (0.001) | (0.001) |
| PastWinner | 0.064*** | 0.064*** | 0.066*** | 0.053*** | 0.028*** |
| | (0.005) | (0.005) | (0.004) | (0.003) | (0.003) |
| ScoreCount | 0.170*** | 0.170*** | 0.175*** | 0.178*** | 0.199*** |
| | (0.001) | (0.001) | (0.001) | (0.001) | (0.001) |
| Constrained Experience (%ile) | | | | | |
| × After | 0.047* | 0.058** | 0.064*** | 0.097*** | 0.079*** |
| | (0.021) | (0.022) | (0.015) | (0.012) | (0.011) |
| × Fake After | | −0.037 | | | |
| | | (0.024) | | | |
| Contest | Fixed | Fixed | Fixed | Fixed | Fixed |
| Individual | Random | Random | Random | Random | Random |
| AIC | −18303.241 | −18297.969 | −35739.401 | −59311.859 | −75407.414 |
| Log Likelihood | 9235.620 | 9233.984 | 18025.701 | 29941.929 | 38316.707 |
| Num. obs. | 18819 | 18819 | 37435 | 73208 | 136989 |
| Num. groups: UserID | 5828 | 5828 | 10614 | 19143 | 33813 |
| Var: UserID (Intercept) | 0.005 | 0.005 | 0.005 | 0.008 | 0.013 |
| Var: Residual | 0.018 | 0.018 | 0.019 | 0.021 | 0.027 |

***$p < 0.001$; **$p < 0.01$; *$p < 0.05$

### 4.1. Examining evaluation signals

We sought to understand possible mechanisms that help to explain why constrained projects facilitate the learning process. The nature of audience evaluation signals has been shown to be important as inputs to the creative process, with past work showing how designers in an online setting, for example, can make use of audience votes on their designs or firm-selected winners of design competitions as signals to improve their own efforts (Riedl & Seidel, 2018). We investigated whether creative constraints narrow the variance of evaluation signals by conducting a Welch two-sample t-test of the vote rating variance between creatively unconstrained ("open" projects) and creatively constrained projects.

We found this difference was significant (t=3.2386, p=0.001**), in the direction expected, with constrained projects having narrower variance. Figure 2 illustrates this difference. The result supports the thesis that if designers are given a "blank slate" and are relatively unconstrained, as in some recommendations for creative outcomes (Amabile, 1983), audiences will be less able to assess relative quality and will therefore have wider evaluation of such work when compared with more constrained design projects. Creative constraints narrow the design space (Beaudouin-Lafon & Mackay, 2009; Boden, 2004), thus also narrowing evaluation signals as well.

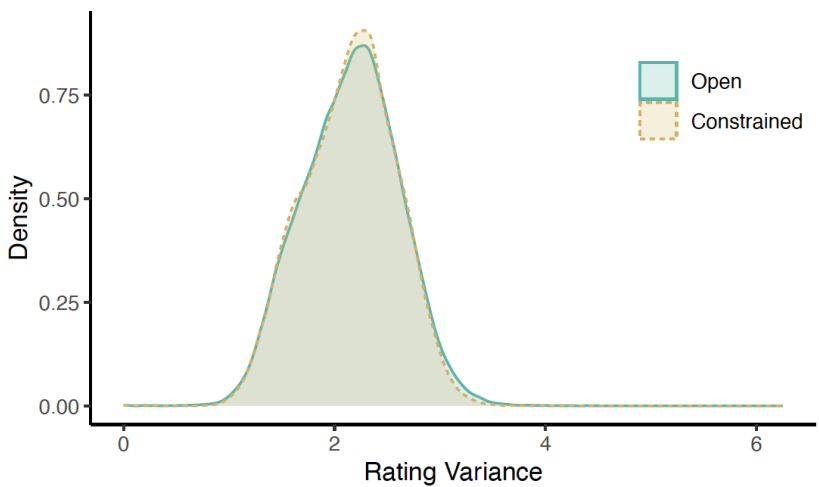

**Figure 2: Comparison of design evaluation ("voter rating") variance between creatively unconstrained and creatively constrained submissions**
Figure notes: Creatively more constrained design projects (in yellow with dotted lines) receive statistically narrower evaluation variance (t=3.2386, p=0.001**) than relatively unconstrained design projects (in teal, with solid lines). Data based on over 140 million votes.

We were also able to determine if there were overall differences in mean score between constrained and unconstrained design projects. While an unconstrained "blank slate" creative endeavor offers freedom to explore (Amabile, 1983), such creative works may be challenging to evaluate by an audience. When compared with a creatively unconstrained design, a design evaluated under creative constraints provides an anchor by which audiences can evaluate the quality (Seifert et al., 2015), which might lead to higher overall evaluation due having a reference point for quality.

A t-test of the mean scores between creatively unconstrained and constrained projects was also significant (t=-21.181, p<0.001**) with constrained projects receiving higher scores. Overall, creatively constrained designs gathered higher scores and less variance from their audience that those that are relatively unconstrained, leading to different audience evaluation signals for designers.

## 5. Discussion

The dominant model for the role of constraints in innovation is that there is an inverted-U relationship to the use of constraints in an innovation task (Acar et al., 2019; Ehls et al., 2020), drawing on a history of studies that demonstrate how constraints can be a means to help initially generate ideas (Stokes, 2005) but can also serve as an impediment that narrows contributions (Smith, 2003). Our extension to this model of the role of constraints provides the finding that the type of constraint is significant and contingent: Creative constraints increase learning how to create successful designs in a design space, but only if intrinsic technical constraints are sufficiently relaxed.

We theorize this dynamic is driven by the need for a broad conceptual design space that is not overly constrained by technical constraints in which to experiment with novelty, coupled with specific-enough creative constraints that focus evaluation criterion for learning. Experimentation is an important part of generating new innovations (Erat & Kavadias, 2008), and when technological constraints are narrow the design space of possibilities itself becomes too constraining. Taking our T-shirt empirical setting example of color choices to the limit, if there were just monochrome options available, one can see how this may limit the degree by which new ideas could be experimented with in creative design, limiting the rate of learning over time. Likewise, in video game design one can consider the very early days when color options and resolutions were extremely limited, such as when games like the simple paddle-tennis monochrome game Pong emerged. Working within such a design space so narrowly bounded by technical constraints can be seen to limit how a designer could imagine new creative designs and learn from their experiments: The range of new Pong-type creative designs was quite limited, in turn limiting creative exploration and learning.

Once technical constraints are reduced to allow a broader range of experimentation, our study found it actually becomes valuable to have some level of creative constraint to help direct effort and promote learning. Our tests of our hypotheses regarding the audience evaluation signals available to designers demonstrated that under creative constraints (such as an imposed theme for a new T-shirt design) the evaluation signals are more focused and higher, which we theorize reflects an increased quality of signal for learning. Making sense of mixed quality signals has been shown to be important in learning within design tasks (Riedl & Seidel, 2018). With increased-quality signals, designers are better able to understand how their design projects are received (whether successful or unsuccessful), and therefore make adjustments that improve their performance in subsequent design submissions. Interestingly, their performance is not limited to further constrained projects, but this improvement is seen in unconstrained ("blank slate") projects as well. Designers can integrate their new understanding of what is valued in the design space for subsequent submissions of either form.

Our study has three main contributions for our understanding of constraints in the context of online innovation communities. First, we demonstrate the effect of constraints tied to learning, expanding from what we know about the effect on creativity in one-off design projects. The repeated process of exploring a design space over multiple submissions is a setting that represents the common iterative nature of much design in practice (Erat & Kavadias, 2008) as well as a central theme for many online innovation efforts (Lakhani & Wolf, 2005; Riedl & Seidel, 2018). While past research has focused on one-off creativity projects across in-person projects (Rosso, 2014), certain online communities, (Ehls et al., 2020) or experimental assessments (Jansson & Smith, 1991), much creative work is far more iterative, involving repeated instances of preparing a submission for audience evaluation and then adapting in advance of the next submission. Indeed, the rise of online platforms for open innovation further increases the opportunities for fast, repeated iterations of designs, where repeated interaction builds sustainable communities (Majchrzak & Malhotra, 2016). Our empirical setting and resulting findings allowed us to extend prior work on one-off project creativity to the realm of individual learning of how to be a more successful contributor over time.

Second, we extend our understanding of the role of constraints to better distinguishing the role of creative versus technical constraints in the context of online innovation communities. While the dominant model has been the inverted-U relationship between constraints and creativity (Acar et al., 2019; Ehls et al., 2020), contingent effects have been less explored, and many past treatments of typologies of constraints (Acar et al., 2019; Kornish & Hutchison-Krupat, 2017) have neglected to examine technical constraints, particularly in the context of online platforms. Online innovation communities have the ability to dramatically increase the number of individuals who can explore a design space, but they are also constrained by the inherent technical limitations. We have provided a parsimonious explanation for how two types of constraints can interact to either limit or promote learning. There are implications for further research within this contribution. While our study looked at creative constraints of either being constrained or unconstrained, future work considering the number of creative constraints imposed could determine the nature of this relationship in more detail, investigating the degree of an inverted-U relationship seen in one-off projects (Acar et al., 2019), and examining this in relation with continuous rather than discrete changes in technical constraints.

Third, our study contributes an understanding of the conceptual spaces that exist alongside the virtual spaces in the process of knowledge creation. Online communities themselves have been conceptualized as "virtual spaces" where the sociality of how people are connected is central to how they create knowledge (Faraj et al., 2017). Virtual spaces are defined by setting constraints on such things as whom is invited to participate, what collaborative and competitive social ties form, and how individuals are organized to learn from each other. The individuals within such virtual spaces are also confronted with conceptual spaces under which they advance new ideas, create new knowledge through collaboration, and seek out possible new innovations (Jin et al., 2021). Design spaces are conceptually bounded by the framing of what are considered acceptable ideas and technologically bounded by the affordances of a platform. The interplay between the constraints inherent within virtual spaces and design spaces together will define an individual's capacity to produce creative output and learn how to improve their efforts, and in turn how the community as a whole develops and evolves over time.

There are two primary managerial implications from our study. First, our study points to the need for managers to re-think the value of purely open "blank slate" challenges, instead they may want to consider a series of constrained, targeted challenges that may best promote learning among the community, leading to improved solutions overall. Our findings may be relevant not only to increasing design skill (Ravasi & Lojacono, 2005) but also to thinking about capability development within organizations more generally, as they explore for organizational and strategic improvements (Levinthal, 1997). Our findings suggest new capabilities can likely to arise out of sequential focused effort, rather than through broad-based unguided exploration. Second, managers should consider that they need to evaluate how technical constraints imposed by the platform are sufficiently relaxed. While much of the discussion on the design of online communities is on matching the community type to the innovation challenge (e.g. Majchrzak & Malhotra, 2016), this research calls attention to the need for managers to consider the effect of technical constraints on the ability of a community to explore and learn over time.

This research highlights opportunities for further studies related to this more complex understanding between constraints, innovation, and learning. While our study looked at creative constraints of either being constrained or unconstrained, future work could investigating the degree of an inverted-U relationship as seen in one-off projects (Acar et al., 2019), and examine this in relation with continuous rather than discrete changes in technical constraints. While past qualitative work has suggested that cycling between more highly-constrained and less-constrained projects can lead to increased motivation for individuals to participate in an online innovation community (Seidel & Langner, 2015); future research on the role of constraints for learning could examine whether there are limits to the degree of constrained work that is beneficial. Lastly, while our study was focused on externally-imposed constraints from the online platform provider, additional research could examine—perhaps using interview or other qualitative methods—whether individual designers learn to impose their own constraints as they design. Past research on creative behavior has found self-imposed constraints to help the creative process (Stokes, 2005). The interplay between external and internal self-imposed constraints may have implications for how individuals best learn when exploring a design space.

# 6. Conclusion

Online innovation communities expand the range of individuals that may contribute to the design and innovation process, and such communities also provide a virtual space of knowledge flows between individuals that can not only help generate new solutions to design problems but also provide the means by which individuals learn to improve their innovative capacity. All design spaces are constrained to some degree, and by considering creative versus technical constraints, we have been able to provide insight into the mechanics of a creative learning process. Our findings show how the benefits or drawbacks of constraints in exploring a design space cannot be addressed without providing clarity about different forms of constraint and how the two interact together. Our study can help researchers and managers understand the contingent nature of creative versus technical constraints in the important work of generating novel innovations and increasing the creative capacity of individuals within communities.